Dear Fellow Quantum Mechanics
Jeremy Bernstein
Abstract: This is a letter of inquiry about the nature of quantum mechanics.

I have been reflecting on the sociology of our little group and as is my wont here are a few notes.

I see our community divided up into various subgroups. I will try to describe them beginning with a small group of elderly but distinguished physicist who either believe that there is no problem with the quantum theory and that the young are wasting their time or that there is a problem and that **they** have solved it. In the former category is Rudolf Peierls and in the latter Phil Anderson. I will begin with Peierls.

In the January 1991 issue of Physics World Peierls published a paper entitled "In defence of 'measurement'". It was one of the last papers he wrote. It was in response to his former pupil John Bell's essay "Against measurement" which he had published in the same journal in August of 1990. Bell, who had died before Peierls' paper was published, had tried to explain some of the difficulties of quantum mechanics. Peierls would have none of it." But I do not agree with John Bell," he wrote," that these problems are very difficult. I think it is easy to give an acceptable account…" In the rest of his short paper this is what he sets out to do. He begins, "In my view the most fundamental statement of quantum mechanics is that the wave function or more generally the density matrix represents our *knowledge* of the system we are trying to describe." Of course the wave function collapses when this knowledge is altered. It is like having an urn filled with known numbers of black and white balls. The probability of next drawing a white ball changes abruptly after say a black ball has been drawn. There is no spooky action at a distance here. Of course the problem in the quantum theory is what is the urn and what are the balls? What is this "system" about which we have knowledge? Is the system just out there somewhere when we don't have knowledge of it? I do not see how anyone can think of this except as another and unacknowledged form of hidden variable theory. This is particularly entertaining here because at the end of his paper Peierls makes a point of rejecting Bohmian mechanics because he says that **it** is a hidden variable theory. In Bohmian mechanics, as we all know, what is hidden is the wave function. The balls are visible.

Anderson concedes that the collapse of the wave function is a problem, but he knows how to solve it "decoherence." I do not know who first introduced this term. (I do know that Schrodinger was the first to introduce the term "entanglement" in a philosophical paper he wrote in the 30's.) But it was Bohm, as far as I know, who first made use of this notion in the sense that we have come to understand it. In his 1951 text he gives a very complete discussion of the Stern-Gerlach experiment. He writes down the entangled wave function for the two spin possibilities. He then squares it to find an expression for the probabilities of the two spin states. This contains cross terms but he argues that in the presence of the magnetic field the phases of these cross terms oscillate so rapidly that the terms effectively vanish and we have the classical expression for the probabilities. This is the decoherence.

Note well that nothing in this mechanism has projected out one of the two terms. That is what the measurement does. Somehow Anderson has persuaded himself that decoherence does this as well. Steve Adler wrote a nice note showing that this is wrong.[1] The Schrödinger equation cannot describe the collapse of the wave function, which is what this is, and **that** is the measurement problem.

Most physicists in my experience fall into either the Alfred E.Newman or the Esther Dyson camps. You will remember that Alfred E.Newman was a regular in Mad magazine.

---

[1] ArXiv quant-ph/0112095

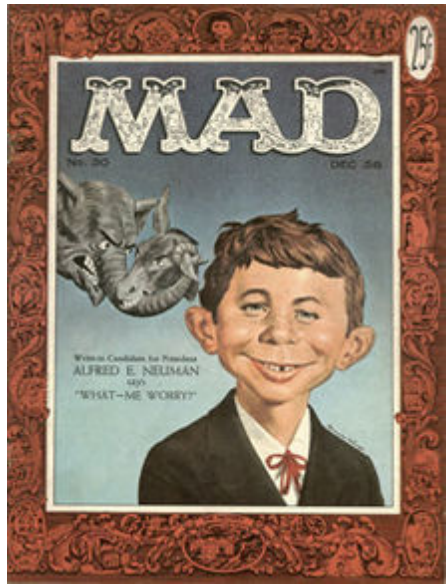

and was noted for saying "What, me worry?" George Dyson is Freeman's son and Esther his somewhat older daughter. Freeman reported to his parents the following conversation between them:

George: I can understand how a boat moves along when you push on the oars. You push the water away and so it makes room for the boat to move along.

Esther: But I can make the boat move along even without understanding it.

Most physicists feel that they can row the quantum boat even without understanding it. This is reflected in the texts-even that of Dirac. There is not a word about a measurement "problem." He briefly discusses a photon "jumping" into a given polarization state after a measurement. That is what happens-period. I am not aware of anything Dirac wrote about these matters, but John Bell once told me that Dirac said to a British physicist that he thought that the book was good but it was missing the first chapter. Speaking of Bell which I did frequently, after a Socratic session with him on the quantum theory one had some sympathy for the Athenians who insisted that Socrates drink the hemlock.

I believe that the revival of interest in such questions can be traced to Bell. There is some irony in this. For Bell this was an avocation. When Bell went to CERN in 1960 is was partly to do elementary particle theory and partly to work on accelerator design. Working on the foundations of quantum theory was not in the job description. This he did in his spare time. But in 1963 Bell was invited to spend a year at Stanford and he felt that he could actually spend time on these matters. It was during this period when he came up with his inequality. As a visitor he felt that he did not have the right to use Stanford money the pay the publication charges for the paper so he published it in the short-lived journal *Physics* which actually paid for articles although the honorarium was about equal to the cost of reprints. Bell might have saved himself the money because until 1969 no one seemed to have paid any attention to it. That year he got a letter from John Clauser at Berkely noting that he -  later there were collaborators- Michael Horne and Abner Shimony from Boston University and Richard Holt of the University of Western Ontario-had produced a generalization of Bell's inequality that might be tested by using polarized light. In 1972 Clauser and Stuart Freedman published the first experimental results and the flood gates opened.

Bell never had the slightest doubt that these experiments would confirm the quantum theory. There was nothing special about the domain in which they were being done, a domain in which all the predictions of the theory were always borne out. But he certainly had no inkling of the reaction to this work. The quantum Buddhists were let loose and are still out there. But highly respectable physicists also got into the act. It was like the old days when people like Bohr, Heisenberg and Einstein discussed the foundations of the theory. For purposes of this little discourse I would like to divide the present collection of workers into two groups. There are the True Believers who think that the quantum theory is a theory of everything-the future, the past-everything. Then there are people who have their doubts. For reasons I will more fully explain I find myself pretty much in the latter camp. It would be odd, I think, if this bi-pedal, carbon-based species located near an uninteresting star should have in a few decades found the Theory of Everything. I had the chance to visit with Schrödinger not too long before his death in his apartment in Vienna. As we were leaving he looked at us intently and said, "There is one thing we have forgotten since the Greeks…modesty!"

The Theory of Everything work that has most impressed me is that of Gell-Mann and Hartle. For reasons I will explain I am not entirely happy with it, but it is impressive. Its ancestral origins are in an obscure paper by Dirac, "The Lagrangian in Quantum Mechanics" which he published in the even more obscure journal *Physikalische Zietschrift der Sowjetunion."* in 1933. Dirac put much of the contents of this paper in subsequent editions of his book which is where Feynman learned about it. It became the subject of his thesis which he did with John Wheeler. The essence of Dirac's paper is summed up in the statement

$$\langle q_T | q_t \rangle \text{ corresponds to } \exp(i \int L(t) dt / \hbar)$$

Here the integral of the Lagrangian L is between the times t and T and no explanation is offered for the meaning of "corresponds to." Dirac then breaks up the time interval into shorter intervals and writes

$$\langle q_t | q_T \rangle = \int \langle q_t | q_m \rangle dq_m \langle q_m | q_{m-1} \rangle dq_{m-1} \ldots \langle q_2 | q_1 \rangle dq_1 \langle q_1 | q_T \rangle.$$

Each one of these factors can also be written in terms of the Lagrangian exponential and hence you have a sum over paths. It is odd that nowhere does Dirac work out any example even for a free particle. He does note that the passage to the classical limit is achieved by letting ℏ tend to zero and noting that the dominant terms in the sum are then over paths that are determined by a stationery action. Feynman does work out several examples in his thesis and after it was published the path integral formalism of quantum theory became an attractive possible alternative formalism. The Gell-Mann- Hartle interpretation is in this spirit.

Suppose, to use an example of Hartle's , we want a "quantum mechanical" description if the Earth's motion around the Sun. We can produce a coarse-grained history by a sequence of projection operators $P_\alpha$ which project

onto successive center of mass positions of the Earth. Then a "history" would be given by

$$C_\alpha = P^k{}_{\alpha_n}(t_k)...P^1{}_{\alpha_1}(t_1).$$

A final state is gotten from an initial state by operating on it with $C_\alpha$. A different choice of projections leads to a different history. One would like to ask how probable is one history compared to another. In general this is not possible unless the two histories decohere, We have seen this in the case of the Stern-Gerlach experiment. If they decohere one can apply the Born rule, which is assumed and not proven, that the probability that a state $|\psi_\alpha\rangle$ is produced from a state

$|\psi\rangle$ is given by $|C_\alpha|\psi\rangle|^2$. One attractive feature of this approach is that it does offer a solution to the measurement problem unless you are churlish enough to insist that an explanation of the Born rule is part of the problem. I would say that it offers the Yogi Berra solution-if you come to a fork in the road-take it. The wave function does not collapse but the other parts which describe alternates to what is actually measured describe other histories-paths not selected. To an Occam's razor kind of guy all those unused paths may seem a little much. As time goes by they grow like rabbits in a field. But this is not what really bothers me about this. It is the past.

      I will state the matter baldly and then explain. I believe that the past is classical while the future is quantum mechanical. Events in the past have happened while events in the future will probably happen. Even some of the founders appeared to think that there was something fishy about trying to describe the past quantum mechanically. Here is Heisenberg in 1930
      "The uncertainty principle refers to the degree of indeterminateness in the possible present knowledge of the simultaneous values of various quantities with which the quantum theory deals; it does not restrict, for example, the exactness of a position measurement alone or a velocity measurement alone. Thus suppose that the velocity of a free electron is precisely known, while the

position is completely unknown. Then the principle states that every subsequent observation of the position will alter the momentum by an unknown and undeterminable amount such that after carrying out the experiment our knowledge of the electronic motion is restricted by the uncertainty relation. This may be expressed in concise and general terms by saying that every experiment destroys some of the knowledge of the system which was obtained by previous experiments."

Then he writes,

"This formulation makes it clear that the uncertainty relation does not refer to the past: if the velocity of the electron is at first known and the position then exactly measured the position for times previous to the measurement may be calculated. Thus for the past times ΔxΔp is smaller than the usual limiting value, but this knowledge of the past is of a purely speculative character, since it can never (because of the unknown change in momentum caused by the position measurement) be used as an initial condition in any calculation of the future progress of the electron and thus cannot be subjected to experimental verification. It is a matter of personal belief whether such a calculation concerning the past history of the electron can be ascribed any physical reality or not."[2]

I think what Heisenberg is saying is that if the initial speed is measured and if there is no subsequent interaction so that this speed is not changed then measuring the position later enables us to retrodict the previous positions. This seems a little wooly to me but the paper of Einstein, R.C.Tolman and B. Podolsky entitled "Knowledge of Past and Future in Quantum Mechanics" published in 1931 is much clearer.[3] This paper, which is in English, was written when Einstein was visiting Caltech. A guess is that the actual writing was done by Podolsky who wrote the EPR paper. Einstein et al present a *gedanken* experiment which purports to show that if past events do not have a quantum mechanical uncertainty then this will lead to a violation of the uncertainty principle for at least some future events. This would seem to be a very profound conclusion. It you believe in a quantum theory of everything then you cannot have a classical past, What I find odd is that this paper does not seem to have inspired much discussion unlike the EPR paper which surely must have caused the destruction of forests to provide the paper on which these discussions were and are being written. I don't find any further discussion by Einstein, nor any by Bohr, nor any by Schrödinger for that matter. Odd.

---

[2] W,Heisenberg, The Phhysical Principles of the Quantum Theory, Dover, New York,1949, p.20
[3] Phys. Rev.37,780-781.

Here is the experiment. Imagine a triangle. A one corner of the base there is a box with a shutter that emits some sort of particle or particles when the shutter opens automatically for a short time. The supposition is that there are so many particles that it happens on one of these openings two particles are emitted. One goes straight across the base to a detector while the other travels around the two sides of the triangle to the detector. They both move at constant speeds such that the one that follows the longer path will arrive later. We have measured these distances. We have also weighed the box before and just after the particles are emitted. This tells us the total energy of the two emitted particles.

Here is what Einstein et all write,

"Let us now assume that the observer at O measures the momentum of the first particle as it approaches along the path SO [The shorter path directly from the source to the observer.], and then measures its time of arrival. Of course the latter observation, made for example with the help of gamma-ray illumination, will change the momentum is some unknown manner. Nevertheless, knowing the momentum of the particle in the past, and hence also its past velocity and energy, it would seem possible to calculate the time when th shutter must have been open from the known time of arrival of the first particle, and to calculate the energy and velocity of the second particle from the known loss in the energy content of the box when the shutter is opened. It would then seem possible to predict beforehand both the energy and the time or arrival of the second particle, a paradoxical result since energy and time are quantities which do not commute in quantum mechanics."

Aside from the fact that time is not generally considered to be represented by an operator, we know what they mean. They go on,

"The explanation of the apparent paradox must lie in the fact that the past momentum of the particle cannot be accurately determined as described. Indeed, we are forced to conclude that there can be no mechanism for measuring the momentum of a particle without changing its value…It is hence to be concluded that the principles of the quantum mechanics must involve an uncertainty in the description of past events which is analogous to the uncertainty for the prediction of future events."

Here they leave the matter without further comment. This result seems to fly in the face of everything we think we know about the past. The Sun **did** rise this morning after all. How does this quantum uncertainty about the past manifest itself? These authors have no comment.

For those of you who may have some interest kindled by this discussion I strongly recommend a paper by Hartle. "Quantum Pasts and the Utility of History"- arXiv:gr-qc/9712001v1/ 2Dec 1997. Hartle accepts the fact that in the quantum theory the future and past are asymmetric. You cannot retrodict the past from present data. He gives the example of the Schrödinger cat. If the feline is observed alive its prior wave function could have been $|ALIVE\rangle$ or some linear combination of $|ALIVE\rangle$ and $|DEAD\rangle$, Incidentally when I visited Schrödinger there was no cat and I was told that he did not like cats. More generally the asymmetry is clear from the expression I gave before for a quantum mechanical history as given by a series of products of projection operators. These do not have inverses. There is no unique past but various ones with different probabilities. When I said to Hartle that the Sun **did** come up this morning with no probabilities he asked how could I be sure? The same set of considerations applies as far as I can tell to Bohmian mechanics. A single Bohmian trajectory is time reversible but from it you cannot construct the probability distribution for the trajectories.

I also recommend a paper by George and Esther Dyson's father. Here is a quote

"I deduce two general conclusions from these thought-experiments. First, statements about the past cannot in general be made in quantum-mechanical language. We can describe a uranium nucleus by a wave-function including an outgoing alpha-particle wave which determines the probability that the nucleus will decay tomorrow. But we cannot describe by means of a wave-function the statement, ``This nucleus decayed yesterday at 9 a.m. Greenwich time''. As a general rule, knowledge about the past can only be expressed in classical terms. My second general conclusion is that the ``role of the observer'' in quantum mechanics is solely to make the distinction between past and future. The role of the observer is not to cause an abrupt ``reduction of the wave-packet'', with the state of the system jumping discontinuously at the instant when it is observed. This picture of the observer interrupting the course of natural events is unnecessary and misleading. What really happens is that the quantum-mechanical description of an event ceases to be meaningful as the observer changes the point of reference from before the event to after it. We do not need a human observer to make quantum mechanics work. All we need is a point of reference, to separate past from future, to separate what has happened from what may happen, to separate facts from probabilities." [4]

Dyson thinks that quantum mechanics applies to "patches" in the universe. One wonders what applies elsewhere and how to find the patches were it does not apply.

I would like to end this letter with a *divinette* which may or may not have anything to do with anything that has gone before. The Planck time is given by $\sqrt{\frac{\hbar G}{c^5}} \sim 5.4 \times 10^{-44}$ sec. If this is a time measured by a clock then this expression violates relativity theory.[5]

Your thoughts?

---

[4] Thought Experiments Dedicated to John Archibald Wheeler, in Science and Ultimate Reality, Cambridge University Press, New York, 2004, 89

[5] Freeman Dyson responded to my divinette with a very pretty argument that shows that the Planck time cannot be measured by clocks.

JB